\documentclass[twocolumn]{revtex4}
\usepackage{latexsym}
\usepackage{amsmath}
\usepackage{amssymb}
\usepackage{amsfonts}
\usepackage{graphicx}
\usepackage{verbatim}

\newcommand{\be}{\begin{equation}}
\newcommand{\ee}{\end{equation}}
\newcommand{\bea}{\begin{eqnarray}}
\newcommand{\eea}{\end{eqnarray}}

\begin{document}

\title{Production and Detection of Axion-like Particles by Interferometry}

\author{H. Tam and Q. Yang}
\affiliation{Department of Physics, University of Florida, Gainesville, FL 32611}

\begin{abstract}
We propose an interferometry experiment for the detection of axion-like particles (ALPs). As in ordinary photon-regeneration (light shining through a wall) experiments, a laser beam traverses a region permeated by a magnetic field, where photons are converted to ALPs via the Primakoff process, resulting in a slight power loss and phase shift. The beam is then combined with a reference beam that originates from the same source. The detection of a change in the output intensity would signal the presence of ALPs (or possibly other particles that couple to the photon in a similar way). Because only one stage of conversion is needed, the signal is of ${\cal O}(g^2_{a\gamma \gamma})$, as opposed to ${\cal O}(g^4_{a\gamma \gamma})$ for photon-regeneration experiments, where $g_{a\gamma \gamma}$ is the coupling between ALPs and photons. This improvement over photon-regeneration is nullified by the presence of shot noise, which however can be reduced by the use of squeezed light, resulting in an improvement in the sensitivity to $g_{a\gamma\gamma}$ over ordinary photon-regeneration experiments by an order of $10^{1/2}$ assuming $10{\rm dB}$ noise suppression. Additionally, our setup can incorporate straightforwardly optical delay lines or Fabry-Perot cavities, boosting the signal by a factor of $n\sim10^3$, where $n$ is the number of times the laser beam is folded. This way, we can constrain $g_{a\gamma\gamma}$ better by yet another factor of $n^{1/2}\sim 10^{1.5}$, as compared to the $n^{1/4}$ boost that would be achieved in photon-regeneration experiments.
\end{abstract}

\maketitle

\section{Introduction}
The exploration of particle physics in the low-energy frontier began with the introduction of the Peccei-Quinn mechanism by Peccei and Quinn to explain the absence of CP violation in the strong interaction \cite{Peccei:1977hh, Peccei:1977ur}.  A by-product of this proposal is a new pseudoscalar particle, known now as the axion \cite{Weinberg:1977ma, Wilczek:1977pj}.  The properties of the axion are essentially characterized by one parameter -- the energy scale at which the PQ symmetry is spontaneously broken, $f_a$. The axion mass and couplings are both inversely proportional to $f_a$.  In the very first axion model, $f_a$ is taken to be the electroweak scale, but this possibility was quickly ruled out by particle and nuclear experiments.  Subsequent models (KSVZ and DFSZ) relax this assumption, and assume $f_a$ could be much larger than originally thought \cite{Kim:1979if, Shifman:1979if,Zhitnitsky:1980tq,Dine:1981rt}.  Using limits from astrophysics (as stellar emission of axions would heat up stars and accelerate their evolution \cite{Dicus:1978fp, Dicus:1979ch,Raffelt:1987yu, Dearborn:1985gp}) and cosmology (avoiding overclosing the universe \cite{Ipser:1983mw, Preskill:1982cy, Abbott:1982af, Dine:1982ah,Sikivie:2006ni}), the value of $f_a$ can be constrained to $10^9 < f_a < 10^{12}$ GeV (or $10^{-15}{\rm GeV^{-1}}<g_{a\gamma\gamma}<10^{-11}{\rm GeV^{-1}}$) which implies that $10^{-6} < m_a < 10^{-3}$ eV.  These constraints in parameter space have additionally been refined by a host of other observations, such as supernova dimming, axion-induced Bremsstrahlung by cosmic rays, distortion to polarizations of gamma ray bursts, photon-photon elastic scattering, axion-induced nuclear moments in cold molecules, etc.  (For details, see, for example, \cite{Moulin:1996vv, Mirizzi:2005ng, Espriu:2010bj, Graham:2011qk, Mena:2011xj}.)  In this mass range, cold axions have the right properties and cosmological abundance to be a substantial fraction of dark matter  \cite{Ipser:1983mw, Preskill:1982cy, Abbott:1982af, Dine:1982ah,Sikivie:2006ni}.

While astrophysical observations and cosmological considerations provide useful constraints on the parameter space, whether axions really exist can only be settled if they are actually detected in the laboratory, and as of today the hypothetical particle remains elusive.  Initially, the likelihood of detecting such weakly interacting particles was deemed low, since a very large $f_a$ implies that axions couple very weakly to ordinary matter (hence are given the name ``invisible axions'').  However, it was pointed out that we can potentially catch glimpses of the elusive particle by exploiting its coupling to two photons, which is given in the Lagrangian by \cite{Sikivie:1983ip,Sikivie:1985yu}
\begin{equation} \label{affdual}
\mathcal{L}_{a\gamma\gamma} = \frac{1}{4}\frac{g_{\gamma}\alpha}{\pi f_a}a\epsilon_{\mu\nu\rho\sigma}F^{\mu\nu}F^{\rho\sigma} = \frac{g_{a\gamma\gamma}}{4} a F \tilde F,
\end{equation}
where $g_{\gamma}$ is a model-dependent coefficient of order unity, $\alpha$ the fine structure constant, $a$ the axion field, $f_a$ the axion decay constant, $F_{\mu\nu}$ the electromagnetic field strength tensor, and $g_{a\gamma\gamma} \equiv g_{\gamma}\alpha/\pi f_a$.  Through this coupling, the axion and photon can therefore mix with each other in a background magnetic field.  It is essentially this principle that underlies the theoretical basis of all existing axion detection experiments. The CERN Axion Solar Telescope and the Tokyo Helioscope, for example, are a realization of the helioscope introduced in \cite{Sikivie:1983ip} and aim to detect axions originating from the Sun, by converting them into X-rays in a strong magnetic field. The photon-axion mixing can also manifest itself in the birefringence and dichroism in the vacuum, resulting in rotation and elliptization of the polarization of light in the presence of a magnetic field. Such signal was already sought in polarimetry experiments such as BFRT \cite{BFRT1} and PVLAS \cite{PVLAS}, and is now among the goals of current experiments such as BMV \cite{BMV} and OSQAR \cite{OSQAR}.

Another type of experiment that makes use of this mixing is photon-regeneration (or so-called ``light shining through a wall'') experiments\cite{Bibber:1987}, in which a small fraction of the photons in a laser beam traverses a region permeated by a magnetic field, where it is converted to axions.  Because of their weak coupling to ordinary matter, the axions then travel essentially unimpeded through a wall, on the other side of which is an identical arrangement of magnets, where some of the axions are induced to convert back to photons, which can be detected.  The primary advantage of photon-regeneration experiments is their greater control over experimental conditions, since the laser beam is prepared in the laboratory, so they do not  have to rely on extraterrestrial axion sources.  The major drawback is that their signal is very weak ($\propto g_{a\gamma\gamma}^{4}$), since two stages of conversion are required.  At the moment, photon regeneration experiments do not have sufficient sensitivity to detect the QCD axion, though they are in principle capable of detecting other particles that couple more strongly to the photon in an analogous manner.  Hence, their primary objective is to detect ``axion-like particles'' (ALPs), rather than axions.

ALPs are predicted to exist generically in string theory \cite{Svrcek:2006yi}.   While pseudoscalar ALPs couple to photons as axions do, scalar ALPs couple to photons via a $aF_{\mu\nu}F^{\mu\nu}$ term in the Lagrangian, so they can be produced by photons whose polarization is perpendicular to the background magnetic field \cite{Maiani:1986md}.  In general, there is no a priori relationship between their mass and couplings of ALPs; hence their parameter space is a lot less constrained compared to axions.

In this paper, we propose a new experimental method based on interferometry to detect ALPs.  A laser beam is split into two beams of equal intensity.  One of them acts as a reference beam, while the other would traverse a region permeated by a magnetic field to induce conversion into ALPs, just as in the first half of photon-regeneration experiments.   However, instead of having a second stage behind a wall where ALPs are converted back to photons, the beam is recombined with the reference beam.  If photon-ALPs conversion has occurred, the beam emerging from the conversion region would have a slightly reduced amplitude and a phase shift relative to the reference beam.  This leads to a change in the combined intensity, which can then be measured by a detector.  Because only one stage of conversion is needed, the signal intensity is proportional to only $g_{a\gamma\gamma}^{2}$, instead of $g_{a\gamma\gamma}^4$ for the photon-regeneration experiment. This, however, does not straightforwardly improve sensitivity to $g_{a\gamma\gamma}$ due to the presence of shot noise in an ordinary light source; we will expound on this later.

In order to avoid having the signal being overwhelmed by the background, the two beams are arranged to traverse paths of different lengths, such that they would be out of phase by $\pi$ at the detector when the magnetic field is switched off.  Thus, without any conversion the two beams would interfere destructively at the detector, and the detection of a flash of light would signal the occurrence of ALPs production.  Unfortunately, at the dark fringe the signal is reduced to a second-order effect ($\mathcal{O}(g_{a\gamma\gamma}^4)$), so it is necessary to modulate the amplitude (or frequency) of the laser by using a Pockels cell.  The presence of the two sidebands in addition to the carrier gives rise to a component in the power output that is of $\mathcal{O}(g_{a\gamma\gamma}^2)$, which can then be isolated and detected by the use of a mixer.

The use of a coherent light source is accompanied by the presence of shot noise.  For an incoming laser beam of $N$ photons, the Heisenberg uncertainty principle implies an fluctuation of $\sqrt{N}$ in the photon number.  This reduces our ability to place a limit on the ALPs-photon coupling: $g_{a\gamma\gamma}\sim B^{-1}L^{-1}N^{-1/4}$, which is the same as that in photon regeneration (where $B$, $L$, and $N$ are the magnetic field, length of conversion region, and number of photons respectively). Fortunately, our design admits a straightforward implementation of light squeezing, which can reduce shot noise by an order of magnitude with current techniques.

Furthermore, by employing optical delay lines or Fabry-Perot cavities, we can enhance the signal by a factor of $n$, where $n$ is the number of times a laser beam is folded. So we can improve our constraint on $g_{a\gamma\gamma}$ by $n^{1/2}\sim 10^{1.5}$. By comparison, the use of an optical delay line or Fabry-Perot cavity in photon-regeneration results in a much weaker improvement of order $n^{1/4}$.

For completeness, we include in our analysis, in addition to ALPs, gravitons, which couple to two photons via the $h_{\mu\nu} T^{\mu\nu}$ coupling in linearized general relativity, where $T^{\mu\nu}$ is the energy-momentum tensor which receives a contribution from the electromagnetic field.  This is known in the literature as the Gertsenshtein Effect \cite{gert:1962, Raffelt:1987im}.  As we will see, the coupling between them is of $\mathcal{O}(G)$, where $G$ is Newton's constant.   With current technologies, it is expected that our proposed experiment clearly does not have the sensitivity to detect gravitons.

We also point out that in recent years there has been a proliferation of hypothesized particles, many of which couple to photons, so they could also potentially be discovered in our proposed experiment.  Some examples include chameleons, massive hidden photons, and light minicharged particles \cite{Holdom:1985ag, Abel:2003ue, Batell:2005wa, Gies:2006ca, Ahlers:2006iz}.  In particular, using results in \cite{Gies:2006ca}, it is straightforward to generalize our analysis to the detection of minicharged particles.

This paper is structured as follows.  In Section II, we review the physical principles underlying the proposed experiment for ALPs and gravitons. In Section III, we propose in detail the experimental design, and discuss how amplitude modulation can boost the signal intensity to $\mathcal{O}(g_{a\gamma\gamma}^2B^2L^2N)$. This is then followed by Section IV, in which we examine the implications of the presence of shot noise; possible ways to enhance the sensitivity of the experiment; and then Section V, in which we discuss methods to help infer the identity of the particle that the laser photons have converted into.

\section{Theoretical Background}
In an external magnetic field, a photon can convert into any particle with a two-photon vertex.  In general, this has two consequences.  First, there is a decrease in the amplitude of the photon, as the newly created particles carry away energy.  Second, if the new particle is massive, a phase shift is introduced.  If the conversion rates for different polarizations are different, the conversion would result in birefringence (ellipticity) and dichroism (rotation of polarization).  We review in this section the theory behind these two effects.

\subsection{Scalar and pseudoscalar ALPs}

\subsubsection{Power loss via the Primakoff Effect}

Photon-axion mixing in a magnetic field is based on the $aF\tilde F$ coupling \eqref{affdual}, where one of the photon legs is a virtual photon in the magnetic field.  If the polarization of the photon is parallel to the magnetic field, the probability of conversion $\eta$ can be obtained from the cross section of this process, which was first done in \cite{Sikivie:1983ip,Sikivie:1985yu} and is given by
\begin{equation}\label{conversion1}
\eta_{\gamma \rightarrow a} = \frac{1}{4 v_a}(g_{a\gamma\gamma} B L)^2\left(\frac{2}{qL}\sin{\left(\frac{qL}{2}\right)}\right)^2,
\end{equation}
where $v_a$ is the velocity of the axion, $B$ the magnetic field, $L$ the length of the conversion region, and $q$ the momentum transfer to the magnet. Since $m_a \ll \omega_{\gamma}\sim {\rm eV}$, the frequency of the laser beam photons, $v_a \sim 1$, $q=m_a^2/2\omega_{\gamma}$. For $L \sim 10 $m, $m_a\sim 10^{-6}{\rm eV}$, this also implies that $qL \sim 10^{-5} \ll 1$. So \eqref{conversion1} can be approximated by
\begin{equation} \label{conversion2}
\eta_{\gamma \rightarrow a} \approx \frac{1}{4}(g_{a\gamma\gamma}B L)^2.
\end{equation}
If we use $B \sim 10 $T, $L \sim 10 $m, and $g_{a\gamma\gamma} \sim 10^{-15}$ GeV$^{-1}$, the probability of photon-axion conversion is of $\mathcal{O}(10^{-26})$.

After the conversion, the amplitude $A$ of the photon is reduced to $A-\delta A$, where
\begin{equation}\label{daaxion}
\delta A_{\gamma \rightarrow a} = \frac{A \eta_{\gamma \rightarrow a}}{2} \approx \frac{g_{a\gamma\gamma}^2 B^2 L^2 A}{8}.
\end{equation}

Equation \eqref{daaxion} is valid when $m_a\ll m_0\equiv\sqrt{2\pi\omega_{\gamma}/L}$, which is about $10^{-4}{\rm eV}$ for given $L$ and $\omega_{\gamma}$. If $m_a$ is larger than $m_0$ the power loss effect decreases rapidly. When $m_a\gg m_0$, $\delta A_{\gamma \rightarrow a} \sim {g_{a\gamma\gamma}^2 B^2 L^2 A}({m_0/ m_a})^4$. To improve the sensitivity one may fill the conversion region with appropriate media to create an effective mass of photons that can be matched with the mass of axions, then the conversion rate restores.

We note that the discussion here is applicable to pseudoscalar ALPs, since they couple to the photon in exactly the same way.  If the photon polarization is instead perpendicular to the magnetic field, the analysis is also valid for scalar ALPs, as they couple to photons via $aFF \sim \vec B \cdot \vec B$ instead.

\subsubsection{Phase lag due to mixing}
When a photon enters a region permeated by a magnetic field, the dispersion relation for the component orthogonal with respect to the magnetic field remains $\omega^2 = k^2$.  However, if axion production occurs, that of the parallel component is modified, and is given by \cite{Maiani:1986md}
\begin{eqnarray} \nonumber
\omega^2 &=& k^2 + \frac{1}{2}\bigg(m_a^2 + g_{a\gamma\gamma}^2 B^2  \\ \label{dispersion}
&& \pm\sqrt{(m_a^2 + g_{a\gamma\gamma}^2 B^2)^2 + 4 g_{a\gamma\gamma}^2 k^2 B^2}\bigg).
\end{eqnarray}

For $B \sim 10 $T and $g_{a\gamma\gamma} \sim 10^{-12} $GeV$^{-1}$,  the value of $g_{a\gamma\gamma}^2 B^2$ is much less than $m_a^2$.  Under this weak mixing assumption, the additional phase acquired $\delta \theta$ (relative to photons that have travelled a distance $L$ but in the absence of a magnetic field) is then approximately\cite{Raffelt:1987im}
\begin{equation}\label{dthetaaxion}
\delta \theta\approx \frac{g_{a\gamma\gamma}^2 B^2 \omega_{\gamma}^2}{m_a^4}({m_a^2L\over 2\omega_{\gamma}}-{\rm sin}({m_a^2L\over 2\omega_{\gamma}})).
\end{equation}
The effect of the phase shift is negligible in comparison with $\delta A/A$ when $m_a\sim 10^{-6}{\rm eV}$. When $m_a\gg m_0$ the effect of the phase shift is comparable or even bigger than $\delta A/A$. However, as we will show in Section III, the signal due to $\delta A/A$ registered by the detector is of first order and the signal due to the phase shift registered by the detector is of second order when one uses amplitude modulation technique. Therefore as far as $\delta A/A\gg (\delta \theta)^2$, the phase shift effect is negligible.

Again, the present analysis on additional phase acquisition applies entirely to pseudoscalar ALPs.  To generalize to scalar ALPs, all we need to do is to  interchanging the parallel and orthogonal components of the photon relative to the magnetic field.  This is expected, as $aF\tilde F \sim \vec E \cdot \vec B$ and $aFF \sim \vec B \cdot \vec B$, and $\vec E_\gamma \perp \vec B_\gamma$.

Even without conversion to axions, the vacuum in the presence of a magnetic field is by itself birefringent, due to loop corrections in QED (the Heisenberg-Euler term: $\frac{\alpha^2}{90m_e^4}[(F_{\mu\nu}F^{\mu\nu})^2 +\frac{7}{4}(F_{\mu\nu}\tilde{F}^{\mu\nu})^2]$) \cite{Heisenberg:1935qt, Schwinger:1951nm, Dobrich:2009kd}. For $B\sim10$T, $L\sim 10$m, $\omega\sim {\rm eV}$, the QED effect of the phase shift is of ${\cal O}(10^{-14})$ so it is registered by the detector of order ${\cal O}(10^{-28})$ which is negligible.

In passing, we point out that we can also use the modified dispersion \eqref{dispersion} to calculate the reduction in the amplitude of the photon, from which we can calculate the power loss obtained in \eqref{conversion2}.

\subsection{Gravitons}

\subsubsection{Power loss via graviton production}
Since the graviton has a two-photon vertex, it can also be created by a photon in an external magnetic field. However, as we will see, photon-graviton conversion is qualitatively different from that between photon and axion.  When photons convert into gravitons, their amplitude is reduced, but this is not accompanied by a phase shift, since the graviton, being massless, moves at the speed of light.  The photon components parallel and perpendicular to the magnetic field convert at equal rates to the two polarization ($+$ and $\times$) of the graviton.  As we will discuss, this qualitative difference can in principle tell us whether gravitons or axions are being produced in the conversion region; for example, we could alternate between modulating the amplitude and phase of the laser beam, which would reveal information about amplitude reduction and phase shift respectively.  We stress that this suggestion is essentially theoretical in nature, as we do not expect our proposed experiment to have the required sensitivity to detect gravitons yet.

In linearized general relativity, the interaction between the graviton and photon is given by
\begin{equation} \label{gravitoncoupling}
S_{h\gamma} = \frac{1}{2}\int d^4x h_{\mu\nu}T^{\mu\nu}_{(\gamma)},
\end{equation}
where $h_{\mu\nu} = g_{\mu\nu} - \eta_{\mu\nu}$ is a small perturbation to the metric, and $T^{\mu\nu}_{(\gamma)}$ is the energy-momentum tensor of the photon, given by
\begin{equation}
T^{(\gamma)}_{\mu\nu} = F_{\mu\rho}{F^{\rho}}_{\nu} - \frac{1}{4} \eta_{\mu\nu}F_{\alpha\beta}F^{\alpha\beta}.
\end{equation}

Without loss of generality, we assume that the graviton and photon propagates along the z axis.  In the transverse-traceless gauge, the $+$ and $\times$ modes of the graviton have respectively the polarization tensors $\epsilon_{11} = -\epsilon_{22} =1$ and $\epsilon_{12}=\epsilon_{21}=1$ (while all other components vanish).  From \eqref{gravitoncoupling}, we see that the graviton-photon coupling $\propto \Sigma_{\alpha}\epsilon^{\alpha}_{ij}(E_i E_j + B_i B_j)$ \cite{Raffelt:1987im}, where $E_i$ and $B_i$ are respectively the electric and magnetic fields, $i=1,2$ and $\alpha = +,\times$.  Hence, the $+$ ($\times$) mode couples only $E_{\perp}$ ($E_{\parallel}$) polarizations, where $\perp$ and $\parallel$ are defined with respect to the plane containing the wave vector of the photon and the external magnetic field.

For simplicity we consider the case where the external magnetic field is perpendicular to the direction of propagation of the photon (and graviton). From the form of the action or the linearized Einstein's equations ($\partial^2 h_{\mu\nu} = -16\pi G T_{\mu\nu}$), we can then compute the conversion probability of photons into gravitons:
\begin{equation} \label{conversiong}
\eta_{\gamma \rightarrow h} = 4\pi G B^2 L^2,
\end{equation}
which is valid for both the $+$ and $\times$ modes.  As expected, \eqref{conversiong} has essentially the same dependence on the magnetic field and length of the conversion region as \eqref{conversion2} ($\propto B^2L^2$).  However, since the Peccei-Quinn scale $f_a$ is far less than the Planck scale $M_{Planck}$, graviton production has a much lower probability than that of axion production. For realistic values of $B$ and $L$ that we used above, the probability is of $\mathcal{O}(10^{-33})$.  So our proposed experiment is not capable of finding the graviton, given existing technologies.

The reduction in photon amplitude corresponding to \eqref{conversiong} is
\begin{equation} \label{dagraviton}
\delta A_{\gamma \rightarrow h} = \frac{A\eta_{\gamma \rightarrow h}}{2} =2\pi GB^2 L^2 A.
\end{equation}

\subsubsection{(The absence of) phase lag due to graviton production}
Unlike the conversion into axions, photon-graviton mixing does not produce a phase lag between the two photon polarizations, so unlike \eqref{dthetaaxion},
\begin{equation} \label{dthetagraviton}
\delta \theta_{\gamma \rightarrow h} = 0.
\end{equation}

This is understandable because the graviton is massless, and therefore moves at the speed of light.  In addition, because both parallel and perpendicular polarizations decay into the the $\times$ and $+$ graviton modes with equal probabilities, birefringence and dichroism do not develop.  As mentioned, this qualitative difference between axion- and graviton-photon mixing in a magnetic field can potentially be utilized to differentiate these particles experimentally.  We also point out that QED birefringence and loop corrections give rise to a slight difference in the production rate of the two graviton polarizations, though they are negligible and therefore ignored here \cite{Ahlers:2008jt}.

\section{Design of Experiment}
\begin{figure}
\includegraphics[bb=0 0 719 539,scale=0.3]{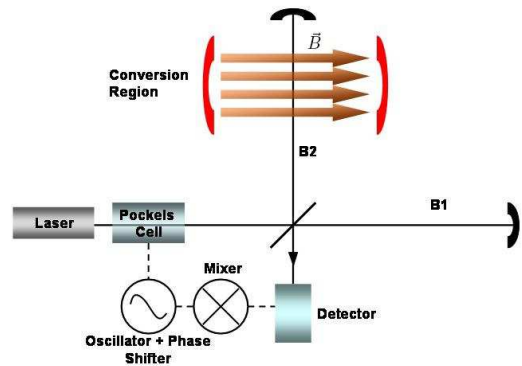}
\caption{Schematic diagram of our proposed experiment.  A laser beam, whose amplitude is modulated by a Pockels cell, is split into two beams of equal intensity ($B_1$ and $B_2$).  The beam $B_2$ (vertical) traverses a region permeated by a magnetic field $\vec B$, where photons convert to axions (and other particles with a two-photon vertex).  It is then recombined at the detector with the beam $B_1$ (horizontal), which acts as a reference.  The two arms are different in length, so that the two beams are out of phase by $\pi$ in the absence of a magnetic field.  A change in intensity registered by the detector would signal the occurrence of a conversion.  To extract the component of the overall signal that is proportional to $g_{a\gamma\gamma}^2$, we mix the output with the oscillator voltage that drives the Pockels cell.}
\label{schematic}
\end{figure}
In our proposed experiment, a laser beam first enters a Pockels cell (with a polarizer behind) to modulate its amplitude (the purpose of the modulation will be explained below).  Subsequently, it is divided by a beamsplitter into two beams (which we label $B_1$ and $B_2$ in Figure~\ref{schematic}) with equal intensity.   $B_2$ is essentially the laser beam used in the first half of the ``shining-light-through-the-wall'' experiment: it passes through a region permeated by a constant magnetic field, where a small fraction of the photons are converted into axions which carry energy away from the beam, according to \eqref{conversion2}.  For simplicity, we will consider here that the carrier of the modulated beam (both $B_1$ and $B_2$) is linearly polarized in the direction of the magnetic field, so our analysis in the previous section applies (For the detection of scalar ALPs, the polarization should be perpendicular to the magnetic field instead).  The two beams are then recombined at the detector, and in the presence of a conversion, the slight amplitude reduction and phase shift would lead to interference, which can be detected.

The length of the path traversed by beam $B_1$ is by design slightly different from that by $B_2$, so that at the detector the two beams would be out of phase by $\pi$ if the magnetic field has been absent.  Operationally, this can be achieved by adjusting one of the path lengths until destructive interference is observed at the detector when the magnetic field is turned off.  Hence, in the absence of the sidebands, the two beams would interfere destructively at the detector.  The purpose for this arrangement is to reduce the background, thereby enhancing the signal-to-noise ratio and minimizing shot noise.

Let the path lengths of the two arms be $L_x$ and $L_y$ (corresponding to beams $B_1$ and $B_2$), and that the state of the laser after passing through the Pockels cell can be described by
\begin{equation}
\vec E_{in} = \vec E_{0} (1+\beta \sin \omega_m t) e^{i\omega t},
\end{equation}
where $\beta$ is a constant, $\vec E_0$ the initial electric field at $t=0$, and $\omega$ is the frequency of the laser.  The amplitude is modulated at a frequency $\omega_m$.  This can be recast as
\begin{equation} \label{Einitial}
\vec E_{in} = \vec E_{0} \left(e^{i\omega t} +\frac{\beta}{2i}e^{i(\omega+\omega_m)t} - \frac{\beta}{2i}e^{i(\omega-\omega_m)t}\right),
\end{equation}
where the first term is referred to as the ``carrier'', and the latter two as ``sidebands''.

The state of the carrier after recombination at the detector is given by
\begin{eqnarray} \nonumber
\vec E_{carrier} &=& -\frac{\vec{E}_0}{2} e^{i(\omega t + 2 kL)} \\
&& \times \bigg[2i \sin k\Delta L - (\frac{\delta A}{A}+i\delta \theta ) e^{-i k\Delta L}\bigg],
\end{eqnarray}
where $k = \omega /c$ is the wavenumber of the laser photons, $A = |\vec E_0|$, $\Delta L = L_x-L_y$ is the length difference between the two arms, and $L = (L_x + L_y)/2$ is the average.  As mentioned, we will choose $k\Delta L=\pi$, so that the detector operates at a dark fringe, in order to eliminate the background signal.  This leads to
\begin{equation}
\vec E_{carrier} = {e^{i(\omega t + 2kL)}\over 2} (\frac{\delta A}{A}+i\delta \theta ) \vec{E}_0.
\end{equation}
Note that without the aid of the sidebands, this would be the entire signal.  While the background is eliminated, the intensity ($\sim \vec E^2$) is of $\mathcal{O}(g_{a\gamma\gamma}^4)$ (for axions).  This loss in sensitivity, as we will see, can be recovered by using the sidebands.

Meanwhile, the sidebands (second and third terms of \eqref{Einitial}) are described by
\begin{eqnarray} \nonumber
\vec E_{\pm} &=& \vec E_0\beta e^{i(\omega t + 2kL)}e^{\pm i(\omega_m t + 2\omega_m L/c)} \\
&& \times \bigg[\sin\frac{\omega_m \Delta L}{c} \mp i(\frac{\delta A}{A}+i\delta \theta) {e^{\mp i \omega_m \Delta L/c}\over 2}\bigg],
\end{eqnarray}
where the subscripts $+$ and $-$ denote respectively the sideband components of frequency $\omega+\omega_m$ and $\omega-\omega_m$.

If we set $\omega_m  \approx \pi c/2\Delta L$, the total electric field at the detector is obtained by adding that of the carrier and sidebands:
\begin{eqnarray} \nonumber
\vec E &=& \vec E_0 e^{i(\omega t+2kL)}\bigg({1\over 2}(\frac{\delta A}{A}+i\delta \theta) \\
&& + \beta \left(2-(\frac{\delta A}{A}+i\delta \theta)\right)\cos\left[\omega_m t + \frac{2\omega_m L}{c}\right]\bigg)
\end{eqnarray}
Note that this particular value of $\omega_m$ is chosen to maximize the signal.  Since $\omega_m \rightarrow n\omega_m$ and $k\Delta L \rightarrow n\pi$ (for $n$ an odd integer) are equally valid choices, the experimenter has much freedom in choosing a suitable value for $\omega_m$ that is experimentally feasible.

Hence, the power $P$ that falls on the detector is
\begin{eqnarray} \nonumber
P &=& P_{in}\bigg\{ \frac{(\delta A/A)^2+\delta \theta ^2}{4}
+ \frac{\beta^2(4-4{\delta A\over A}+{\delta A^2\over A^2}+\delta\theta ^2)}{2} \\ \nonumber
&+& \beta (2{\delta A\over A}- {\delta A^2\over A^2}+{\delta\theta^2\over 2})\cos \left[\omega_m \left(t + \frac{2 L}{c}\right)\right] \\
&+& \frac{\beta^2(4-4 \frac{\delta A}{A}+ \frac{\delta A^2}{A^2}+\delta \theta ^2)}{2}\cos \left[2\omega_m \left( t + \frac{2L}{c}\right)\right]\bigg\}\nonumber\\.
\end{eqnarray}

Thus the power has a dc component (first line), and two ac components with frequencies $\omega_m$ and $2\omega_m$.  If we multiply this with the oscillator voltage that drives the Pockels cell (plus an appropriate phase shift) via a mixer, we can extract the component of frequency $\omega_m$.  Neglecting the second-order contributions, the time-averaged output power of the mixer is given by
\begin{eqnarray}
P_{out} &=& \frac{1}{T} \int_T 2 P_{in}\beta \mathcal{G} \left(\frac{\delta A}{A}\right) \cos^2 \left(\omega_m t\right) \\
&=& \frac{P_{in} \beta\mathcal{G}\delta A}{A}
\end{eqnarray}
where $\mathcal{G}$ is the gain of the detector and $T$ is taken to be sufficiently long to ensure that the time-averaging is accurate.  Hence, the output signal is proportional to $g_{a\gamma\gamma}^2$ for axions and $G$ for gravitons.

In this analysis we choose to modulate the amplitude, rather than the phase, of the photons so the result will not be spoiled by the QED effect. In principle, we could instead modulate the phase, in which case the change in intensity registered by the detector would be primarily a consequence of the phase shift instead of the amplitude reduction.  The corresponding analysis is highly analogous and will not be repeated here.  The major difference is that the coefficients for the sidebands in \eqref{Einitial}, $\beta/2i$, are replaced approximately by $J_1(\beta)$,  the first-order Bessel function of the first kind (higher harmonics now are also present, but are negligible).  Since $J_1(\beta)$ are real, our earlier analysis would work if $\delta A/A$ is replaced by $i\delta\theta$, which is purely imaginary.  This can be implemented by manipulating polarizers adjacent to the Pockels cell.
Thus by switching between phase and amplitude modulation, we can infer information on both the amplitude reduction and phase shift.  This is one conceivable way of identifying the particles that the photons have converted into. For the experiments mainly interested in measuring the QED effect, the phase modulation should be employed.

\section{Sensitivity}
Despite the improvement in signal size, the use of interferometers is inevitably accompanied by the presence of shot noise, which is a manifestation of the granular nature of the coherent state of photons in the laser beam. This limits the resolution of the interferometer therefore reducing the sensitivity to $g_{a\gamma\gamma}$ in our set up.

For a laser beam consisting of $N$ incoming photons, we expect the shot noise in our setup to have a magnitude of $\sqrt{N}$ due to Poisson statistics.  The signal-to-noise ratio is thus reduced to $(g_{a\gamma\gamma}BL)^2 N/\sqrt{N}$.  In the case of a non-detection, this allows us to constrain the axion-photon coupling to $g_{a\gamma\gamma, max}< (BL)^{-1}N^{-1/4}$, which is what can be achieved by conventional photon-regeneration experiments. (In their case, the signal is much smaller, of $\mathcal{O}(g_{a\gamma\gamma}^4 N)$, so dark count rate can be a problem.)

Our setup admits a straightforward implementation of squeezed light using standard optical techniques, which can help reduce shot noise. Using interferometry, in principle, is a different realization of the polarimetry experiment that measures birefringence and dichroism. However, in the polarimetry the dominant noise is the intrinsic birefringence of the optical devices. In interferometry, the intrinsic noises is dominated by the photon counting error (shot noise). Shot noise can be viewed as the beating of the input laser with the vacuum fluctuations entering the other side of the beam splitter. The conception of reducing shot noise by injecting squeezed light is first suggested by \cite{caves}. Let us give a brief summary in the following. The coherent state $|\alpha>$ is described by the unitary displacement operator: $|\alpha>=D(\alpha)|0>=exp(\alpha a^{\dagger}-\alpha ^*a)|0>$, where $a^{\dagger}$ and $a$ are creation and annihilation operators of photons with frequency $\omega$ and $\alpha$ is a complex number. The photon number operator is $N=a^{\dagger}a$ and one finds: $\bar N=|\alpha|^2$, $\Delta N=|\alpha|$ for the coherent state. A squeezed state is described as $|\alpha,\zeta>=D(\alpha)S(\zeta)|0>$, where $\zeta=re^{i\theta}$ is a complex number and $S=exp[1/2(\zeta^*a^2-\zeta(a^{\dagger})^2)]$. For the squeezed state one finds: $\bar N=|\alpha|^2+sinh^2r$ and $(\Delta N)^2=|\alpha cosh r-\alpha^*e^{i\theta}sinh r|^2+2cosh^2rsinh^2r$. Let mode $1^+$ denote electromagnetic field incident from the laser side of the beam splitter and mode $2^+$ denote electromagnetic field incident from the other side of beam splitter. By using an ordinary laser $|\alpha,0>$ in one side of the beam splitter and injecting squeezed light $|0,\zeta>$ from the other side of beam splitter we have the state: $|\phi>=S_2(\zeta)D_1(\alpha)|0>$. The photons received by an ideal photo-detector in one output port then have the property: $\bar N=\alpha^2 sin^2(\phi/2)+cos^2(\phi/2)sinh^2r$ and $\Delta N^2=\alpha^2sin^4(\phi/2)+2cos^4(\phi/2)cosh^2rsinh^2r+sin^2(\phi/2)cos^2(\phi/2)(\alpha^2e^{-2r}+sinh^2r)$, where $\phi$ is the phase difference between the two arms of the interferometer. We see that if one operates near a dark fringe, $\bar N=sinh^2r$ and $(\Delta N)^2=2cosh^2rsinh^2r$ which can be arbitrarily small in theory. Implementations of squeezed light together with using power recycling and sidebands are demonstrated by \cite{squeezwithcave} and later a $10$dB shot noise reduction is achieved \cite{squeezwithcave2}. A $10{\rm dB}$ suppression of shot noise can result in a $10^{1/2}$ improvement of the constraint to $g_{a\gamma\gamma}$.

To further boost the sensitivity, we can incorporate in our setup optical delay lines or Fabry-Perot cavities to enhance the signal by a factor of $n$, where $n$ is the number of times the laser beam is folded. The resultant improvement in our ability to constrain $g_{a\gamma\gamma}$ is of order $n^{1/2}\sim 10^{1.5}$ v.s. $n^{1/4}\sim 10^{0.75}$ in photon regeneration experiment. Combined, the use of squeezed light and optical delay lines results in a gain in the sensitivity to $g_{a\gamma\gamma}$ of $10^{2}$ over a simple photon regeneration experiment.

If we use $n\sim 10^3$, $B\sim 10{\rm T}$, $L\sim 10{\rm m}$ with a $10$W $(\lambda=1\mu{\rm m})$ laser, after $240$ hours running, the experiment can exclude ALPs with $g_{a\gamma\gamma}>2.8\times 10^{-10}{\rm GeV^{-1} }$ to $5\sigma$ significance. If one also employs squeezed-light laser which improves signal-to-noise ratio by $10$dB with similar setup, the exclusion limit can reach $g_{a\gamma\gamma}\sim10^{-11}{\rm GeV^{-1}}$.

Even without employing the squeezed light technique, a interferometer experiment boosted by a Fabry-Perot cavity can achieve similar sensitivity as that in the resonantly-enhanced axion-photon regeneration experiment (a purely laboratory experiment probing axion-photon coupling at a level competitive with or superior to limits from stellar evolution or solar axion searches) \cite{resregeneration}. In practice, the interferometer experiment does not require sophisticated locking and alignment techniques for two cavities which is essential to the resonantly-enhanced axion-photon regeneration experiment. In addition, the interferometer experiment has doubled photon to axion conversion length with same amount of magnets so the signal is boosted by a factor of four.

\section{Discussions and Conclusions}

Now is an exciting time for particle physics.  In the high-energy frontier, the LHC has finally begun operation, providing us with unprecedented access to physics at the TeV scale.  In the low-energy frontier, a large number of experiments worldwide (e.g. ADMX \cite{ADMX}, CAST \cite{CAST}, PVLAS \cite{PVLAS}, GammeV \cite{GammeV}, CARRACK \cite{CARRACK}, ALPS(at DESY) \cite{ALPS}, OSQAR(at CERN) \cite{OSQAR}, etc.) are currently actively searching for new physics at the sub-eV scale, with a particular focus on discovering light scalars, most notably the QCD axion, and also ALPs, hidden photons, and chameleons, among other exotic particles.  For a good summary of existing and future experiments, the reader is referred to \cite{Jaeckel:2010ni}.

The exploration of new physics at the low-energy frontier is a well-motivated enterprise.   After all, more than thirty years have passed and the axion remains the most attractive solution to the strong CP problem.  Even more remarkably, unbeknownst originally to the pioneers in axion physics, the properties of their new creation turn out to match precisely with those of dark matter (for a summary of evidence that favors axionic dark matter, see \cite{Sikivie:2010yn}).   With the realization that ALPs exist abundantly in string theory \cite{Svrcek:2006yi}, there are thus ample reasons to believe that new physics might lurk at the sub-eV scale, waiting to be discovered.

In this paper, we propose a new method of ALPs detection based on interferometry.  A laser source is split into two beams, where one is exposed to a magnetic field permeating a confined region, within which photon-axion conversion occurs.  This results in a phase shift and reduction in amplitude, which can be made manifest if the beam is then recombined and made to interfere with the other, which acts as a reference. In order to avoid the signal being overwhelmed by the background, it is necessary to have the detector operate at a dark fringe.  Unfortunately this also reduces the signal to a second-order effect ($\mathcal{O}(g_{a\gamma\gamma}^4)$).  This reduction can be nullified by modulating the photon amplitude, and mixing the output signal with the oscillator voltage that drives the Pockels cell.

While we have as our principal aim the detection of ALPs, our design is theoretically applicable to any particle with a two photon vertex, so that mixing in the presence of an external magnetic field is permitted.  Given the possibility that more than one such particle exists, it is important to identify what the photons have converted into.  We suggest two methods that can help shed light on this issue.  First, we could repeat the experiment by modulating the phase instead of the amplitude of the laser, as this would reveal information about the phase shift as well.  Secondly, scalar and pseudoscalar ALPs can be distinguished by modifying the polarization of the laser.  Conversion can only occur if the polarization is parallel (perpendicular) to the external magnetic field for pseudoscalar (scalar) ALPs.

\section{Acknowledgement}

We thank Guido Mueller and Pierre Sikivie for useful conversations.  This work was supported in part by the U.S. Department of Energy under contract DE-FG02-97ER41029.

\end{document}